\documentclass[reprint,showpacs,amsmath,amssymb,aps]{revtex4-1}
\usepackage{graphicx}
\usepackage{dcolumn}
\usepackage{bm}
\usepackage{hyperref}

\begin{document}

\preprint{} 

\title{Dimensional Effects on the Density of States in Systems with Quasi-Relativistic Dispersion Relations and Potential Wells}

\author{A. Pokraka}
 \affiliation{Department of Physics, University of Alberta, Centennial Centre for Interdisciplinary Science, Edmonton, Alberta, Canada T6G 2E1.}
 \email{pokraka@ualberta.ca}

\author{R. Dick}
 \affiliation{Department of Physics and Engineering Physics, University of Saskatchewan, 116 Science Place, Saskatoon, Saskatchewan, Canada SK S7N 5E2.}
 \email{rainer.dick@usask.ca}

\begin{abstract}
Motivated by the recent discoveries of materials with quasi-relativistic dispersion relations, 
we determine densities of states in materials with low dimensional substructures and relativistic dispersion relations. 
We find that these dimensionally hybrid systems yield quasi-relativistic densities of states that are a superposition of the corresponding two- and three-dimensional densities of states.
\end{abstract}


\maketitle

\section{\label{sec:level1}Introduction}

Quantum mechanics and electrodynamics in one or two dimensions are commonly employed to describe the behaviour of particles or quasi-particles (such as phonons) in low-dimensional substructures like quantum wires or quantum wells. 
However while particle wave functions may be compressed and squeezed, they will always extend beyond the boundaries of attractive potential wells.
Therefore it is of interest to develop analytic models to study the transition between two-dimensional and three-dimensional behavior of particles in these systems. 

The transition between two-dimensional and three-dimensional behavior of particles in the presence of low-dimensional substructure has been examined for non-relativistic dispersion relations both for quantum wells \cite{1} and for substructures where particles propagate with different effective mass (see e.g. \cite{2} and references there).
The quasi-relativistic case with different effective mass in a substructure has been studied in Ref. \cite{3}.
All these systems exhibit inter-dimensional behavior in the local density of states (DOS) in the sense that energy and length scales can be identified in which the density of states approaches the well-known two-dimensional or three-dimensional limits. 

In this paper we extend these studies to the case of particles with quasi-relativistic dispersion relations in the presence of quantum wells. 
Specifically, we determine the DOS of charged quasi-relativistic bosons in the presence of a thin quantum well.

The DOS counts the number of states per unit volume at a given energy and position, and is important for understanding various properties of materials - in particular, for estimating the availability of carriers for charge and heat transport. It also impacts scattering and absorption 
in materials and signifies band gaps. 
Confining potentials change the local density of states in a way that manifests the interplay between three-dimensional and low-dimensional effects.
For example, consider a thin interface in a three-dimensional bulk material, modeled as a $\delta$ well potential. 
In Schr\"{o}dinger theory, the DOS at the interface is \cite{1} 
\begin{multline} \label{SE DOS}
\rho(E,z_{0})=\kappa\rho_{n=2}(E+(\hbar^{2}\kappa^{2}/2m)) 
\\
+\rho_{n=3}(E)\left[1-\frac{\hbar\kappa}{\sqrt{2mE}}\arctan\left(\frac{\sqrt{2mE}}{\hbar\kappa}\right)\right]
\end{multline}
 where $\mathcal{W} = q l \Phi_0$ parameterizes the effective thickness and depth of the well, $\kappa = m \mathcal{W}/\hbar^2$ is the inverse penetration depth, and $\rho_{n=2}(E)=\Theta(E)\frac{m}{\pi\hbar^{2}}$ and $\rho_{n=3}(E)=\Theta(E)\frac{m}{\pi^{2}\hbar^{3}}\;\sqrt{2mE}$ are the well known 
two- and three-dimensional DOS with two spin states. 
The DOS at the interface is a superposition of the two- and three-dimensional DOS. 
The factor $1-\hbar\kappa/\sqrt{2mE}\arctan(\sqrt{2mE}/(\hbar\kappa))$ smoothly turns on the three-dimensional contribution to the DOS. 
Equation \eqref{SE DOS} confirms the intuitive assumption that bound states in the well contribute to a two-dimensional DOS which is made dimensionally correct in three dimensions through scaling by the inverse penetration depth, $\kappa$. 
Note also that $B=\hbar^2\kappa^2/2m$ is the binding energy of particles in the well, i.e. their energy is \begin{equation} \label{eq:Enonrel} 
E=\frac{\hbar^2k_\|^2}{2m}-\frac{\hbar^2\kappa^2}{2m}=\frac{\hbar^2k_\|^2}{2m}-\frac{m\mathcal{W}^2}{2\hbar^2},
\end{equation}
where $\hbar k_\|$ is the momentum parallel to the potential well.
The argument of the two-dimensional contribution to the density of states in \eqref{SE DOS} is therefore the kinetic energy of particles in the well.

Research into materials with quasi-relativistic dispersion relations has exploded since the discovery of graphene in 2004 \cite{4}. 
These materials include two and three-dimensional Dirac semi-metals \cite{5,6}, topological insulators \cite{7,8}, topological Dirac semi-metals \cite{9,10} and superfluid phases of $^3$He \cite{11}. 
All of these materials have low energy quasiparticle excitations described by Dirac Hamiltonians and thus linear dispersion relations centred around Dirac points in momentum space \cite{11}. 
In particular, these materials posses novel electronic properties and thus have promising applications in spintronics and quantum computing. 
 
In this paper we study two related quasi-relativistic inter-dimensional systems, each implemented through a confining electrostatic potential in the Klein-Gordon equation. 
In spite of the apparent relevance of quasi-relativistic dispersion relations for modern materials science, there has not been much recent activity on low-dimensional potentials with quasi-relativistic wave equations. Sveshnikov and Silaev have analyzed spectra and wave functions in $(1+1)$-dimensional systems with imaginary spatial translations \cite{12}, and Ananchenko \textit{et al.} have studied bound states in a $(2+1)$-dimensional Dirac equation with a spherically symmetric potential well \cite{13}.
We are considering Klein-Gordon type systems in $3+1$ dimensions where the low-dimensional substructure is a planar quantum well at $z=z_0$.
We are also focusing on the local density of states since this provides an excellent probe for the transition between two-dimensional and three-dimensional behavior \cite{1,2,3}.
In Section \ref{sec:level2}, we calculate the inter-dimensional Klein-Gordon Green's function for a delta well potential to first order in the particle charge, $q$. 
Then using the relativistic generalization of the well known relation between the imaginary part of the Green's function and the DOS, we calculate the inter-dimensional DOS at the well interface. 
In Section \ref{sec:level3} we calculate, numerically to full order in q, the inter-dimensional DOS inside shallow finite square wells. 
In both sections 2 and 3 we examine under what circumstances we find two and three dimensional behaviour in our model systems. 
Our conclusions are presented in Section \ref{sec:level4} and some mathematical details are collected in appendices A and B.

\section{\label{sec:level2}Delta Well Potentials}

The relativistic generalization of the relation between the imaginary part of the Green's function and the DOS is \cite{3} 
\begin{equation}	\label{Rel Green's to DOS}
\rho(E,z) - \bar{\rho}(\bar{E}, z) = \frac{2E}{\pi \hbar^2 c^2} \Im \langle \boldsymbol{x}_\parallel, z | G(E) | \boldsymbol{x}_\parallel, z \rangle
\end{equation}
where $\bar{\rho}$ is the anti-particle DOS, $\bar{E} = -E$ is the anti-particle energy and $\Im$ signifies the imaginary part of a complex number. 
We calculate the Green's function for the quasi-relativistic
quantum well and
then use \eqref{Rel Green's to DOS} to determine the inter-dimensional DOS.

We consider a three-dimensional bulk material with a thin interface. Inside the interface, the electrostatic potential differs by a constant factor $\Phi_0$. 
We approximate the interface as a delta well, centred at $z=z_0$, so that the electromagnetic potential is given by $A^\mu = (\Phi(\boldsymbol{x})/c, \boldsymbol{0})$ 
where $\Phi(\boldsymbol{x}) = -l \Phi_0 \delta(z - z_0)$. The parameter, $l$, has dimensions of length (needed to make the potential dimensionally 
correct as the delta function has units of inverse length) and intuitively parameterizes the thickness of the well. 

The equation of motion for coupling of charged Klein-Gordon and electromagnetic fields is
\begin{equation} \label{KG eq}
\left( \left(\partial_\mu - i \frac{q}{\hbar} A_\mu\right) \left(\partial^\mu - i \frac{q}{\hbar} A^\mu\right) 
 - \frac{m^2c^2}{\hbar^2} \right) \phi (x^0, \boldsymbol{x} ) = 0.
\end{equation}
Substituting $A^\mu$ and keeping only leading order terms in $q$ yields the equation of motion
\begin{equation} \label{int dim KG}
\bigg( \mathcal{\partial_{\mu}\partial^{\mu}}-\frac{m^{2}c^{2}}{\hbar^{2}}+2i\frac{q\Phi_0l}{c\hbar}\delta(z-z_{0})\partial_{0} \bigg) \phi(x^0, \boldsymbol{x})=0
\end{equation}
where we use the convention $\eta{}_{_{00}}=-1$ for the Minkowski
metric. 

Since the retarded Green's function is related to the density of states
by equation \eqref{Rel Green's to DOS}, we look for the $x$-representation of the Green's function, $\left\langle x|G|x^{\prime}\right\rangle $,
that satisfies
\begin{multline} \label{KG Greens}
\left\{ -\partial_{0}^{2}+\nabla^{2}-\frac{m^{2}c^{2}}{\hbar^{2}}+2i\frac{q\Phi_0l}{c\hbar}\delta(z-z_{0})\partial_{0}\right\} \left\langle x|G|x^{\prime}\right\rangle \\
=-\delta(z-z^{\prime})\delta(x^{0}-x^{\prime0})\delta(\boldsymbol{x}_{\parallel}-\boldsymbol{x}_{\parallel}^{\prime}).
\end{multline}

The solution to equation \eqref{KG Greens}, detailed in Appendix A, has the form of a Hankel transformation
\begin{multline} \label{KG Greens xrep}
\left\langle z,\boldsymbol{x}_{\parallel}|G(E)|\boldsymbol{x}_{\parallel}^{\prime},z^{\prime}\right\rangle |_{E=\hbar ck^{0}} =
  \\
  \int\frac{dk_{\parallel}}{2\pi}J_{0} \left( k_{\parallel}\left|x_{\parallel}-x_{\parallel}^{\prime}\right| \right) 
k_{\parallel}\left\langle z|G(E,\boldsymbol{k}_{\parallel})|z^{\prime}\right\rangle 
\end{multline}
 where 
\begin{multline} \label{eq:7}
\left\langle z|G(E,\boldsymbol{k}_{\parallel})|z^{\prime}\right\rangle |_{E=\hbar ck^{0}} = \frac{\Theta\left((k^{0})^{2}
-k_{\parallel}^{2}-\frac{m^{2}c^{2}}{\hbar^{2}}\right)}{2\sqrt{(k^{0})^{2}-k_{\parallel}^{2}-\frac{m^{2}c^{2}}{\hbar^{2}}}} 
\\
\times i \left\{ \exp\left[i\sqrt{(k^{0})^{2}-k_{\parallel}^{2}-\frac{m^{2}c^{2}}{\hbar^{2}}}\left|z-z^{\prime}\right|\right] \right. 
+\frac{ql\Phi_0}{c\hbar}k^{0}
 \\
\times \left. \frac{i\exp\left[i\sqrt{(k^{0})^{2}-k_{\parallel}^{2}-\frac{m^{2}c^{2}}{\hbar^{2}}}\left(\left|z-z_{0}\right|
+\left|z^{\prime}-z_{0}\right|\right)\right]}{\sqrt{(k^{0})^{2}-k_{\parallel}^{2}-\frac{m^{2}c^{2}}{\hbar^{2}}}-i\frac{ql\Phi_0}{c\hbar} k^{0}}\right\} 
 \\
+ \frac{\Theta\left(\frac{m^{2}c^{2}}{\hbar^{2}}+k_{\parallel}^{2}-(k^{0})^{2}\right)}{2\sqrt{\frac{m^{2}c^{2}}{\hbar^{2}}+k_{\parallel}^{2}-(k^{0})^{2}}}
\\
\times \left\{ \exp\left[-\sqrt{\frac{m^{2}c^{2}}{\hbar^{2}}+k_{\parallel}^{2}-(k^{0})^{2}}\left|z-z^{\prime}\right|\right]\right. + \frac{ql\Phi_0}{c\hbar}k^{0}
 \\
\times \left.\frac{\exp\left[-\sqrt{\frac{m^{2}c^{2}}{\hbar^{2}}+k_{\parallel}^{2}-(k^{0})^{2}}
\left(\left|z-z_{0}\right|+\left|z^{\prime}-z_{0}\right|\right)\right]}{\sqrt{\frac{m^{2}c^{2}}{\hbar^{2}}
+k_{\parallel}^{2}-(k^{0})^{2}}-\frac{ql\Phi_0}{c\hbar} k^{0}-i\epsilon}\right\} .
\end{multline}

Equation \eqref{eq:7} is the Greens function for a quasi-relativistic boson in the presence of a thin quantum well, valid for arbitrary 
position and energy. Consistently, in the limit $l, \Phi_0 \rightarrow 0$ equation \eqref{eq:7} reproduces the free Klein-Gordon Green's function.  
It is
not easy to identify a two-dimensional limit from this Green's function and therefore we examine the DOS in the presence of the well in order to 
explore this question further. 

Substituting \eqref{KG Greens xrep} into \eqref{Rel Green's to DOS} yields the inter-dimensional DOS. However we are mostly interested in the 
DOS at (or near) the interface because this is where we expect the most prominent inter-dimensional effects. The density of states at the 
interface ($z=z^{\prime}=z_{0}$) is
\begin{multline} \label{KG DOS}
\rho(E,z_{0})-\overline{\rho}(\overline{E},z_{0}) = 
\\
\Theta(E)\rho_{n=2}(\tilde{E})\left(\frac{1}{1+\frac{\mathcal{W}^{2}}{\hbar^{2}c^{2}}}\right)\frac{\mathcal{W}\tilde{E}}{\hbar^{2}c^{2}}+\rho_{n=3}(E)
\\
\times\left[1-\frac{\mathcal{W}E}{\hbar c\;\sqrt{E^{2}-m^{2}c^{4}}}\arctan\left(\frac{\hbar c\;\sqrt{E^{2}-m^{2}c^{4}}}{\mathcal{W}E}\right)\right]
\end{multline}
 where we used $\mathcal{W}=ql\Phi_{0}$ as in the non-relativistic delta well, the energy variable in the two-dimensional contribution 
is $\tilde{E}=E\;\sqrt{1+(\mathcal{W}/\hbar c)^2}$, and the two- and three-dimensional densities of states are
$\rho_{n=2}(E)=\Theta(E^{2}-m^{2}c^{4})\frac{E}{2\pi(\hbar c)^{2}}$
and $\rho_{n=3}(E)=\Theta(E^{2}-m^{2}c^{4})\frac{E\;\sqrt{E^{2}-m^{2}c^{4}}}{2\pi^2(\hbar c)^{3}}$. Here, like in equation \eqref{SE DOS}, $\mathcal{W}$ 
parametrizes the effective thickness and depth of the well.

The key features of \eqref{KG DOS} are analogous to those of \eqref{SE DOS}. Equation \eqref{KG DOS} is a superposition of the standard  
two- and three-dimensional relativistic DOS, Fig. 1. The states perpendicular to the interface are exponentially suppressed and contribute 
a term proportional to the two-dimensional DOS. 
The argument of the two dimensional DOS is scaled by a factor $\sqrt{1+(\mathcal{W}/\hbar c)^2}$ due to the presence of the well. 
We can understand the appearance of this factor in the two-dimensional contributions to \eqref{KG DOS} by noting that the bound energy eigenstates
in the relativistic delta well are
\begin{equation}\label{eq:relbound}
\psi_{\kappa,\bm{k}_\|}(\bm{x}_\|,z)=\frac{\sqrt{\kappa}}{2\pi}\exp(i\bm{k}_\|\cdot\bm{x}_\|-\kappa |z|)
\end{equation}
with the transverse damping factor
\begin{equation}\label{eq:relkappa}
\kappa=\frac{\mathcal{W}E}{\hbar^2 c^2}
\end{equation}
and energy
\begin{equation}\label{eq:relEbound}
E=c\sqrt{\frac{mc^2+\hbar^2k_\|^2}{1+(\mathcal{W}/\hbar c)^2}},
\end{equation}
i.e. the argument of the two-dimensional contribution is actually the rest energy plus kinetic energy for motion of the bound states along the well,
\begin{equation}
\tilde{E}=c\sqrt{mc^2+\hbar^2k_\|^2},
\end{equation}
and the dimensional factor that converts the relativistic two-dimensional DOS into the contribution from the well states to the three-dimensional 
DOS is $\kappa/\sqrt{1+(\mathcal{W}/\hbar c)^2}$.

The factor $1-\frac{\mathcal{W}E}{\hbar c\;\sqrt{E^{2}-m^{2}c^{4}}}\arctan\left(\frac{\hbar c\;\sqrt{E^{2}-m^{2}c^{4}}}{\mathcal{W}E}\right)$ 
smoothly turns the 
three-dimensional contribution on; in other words it is responsible for the transition between two- and three-dimensional behavior. 

\begin{figure}[h]
		\centering
		\includegraphics[scale=.6]{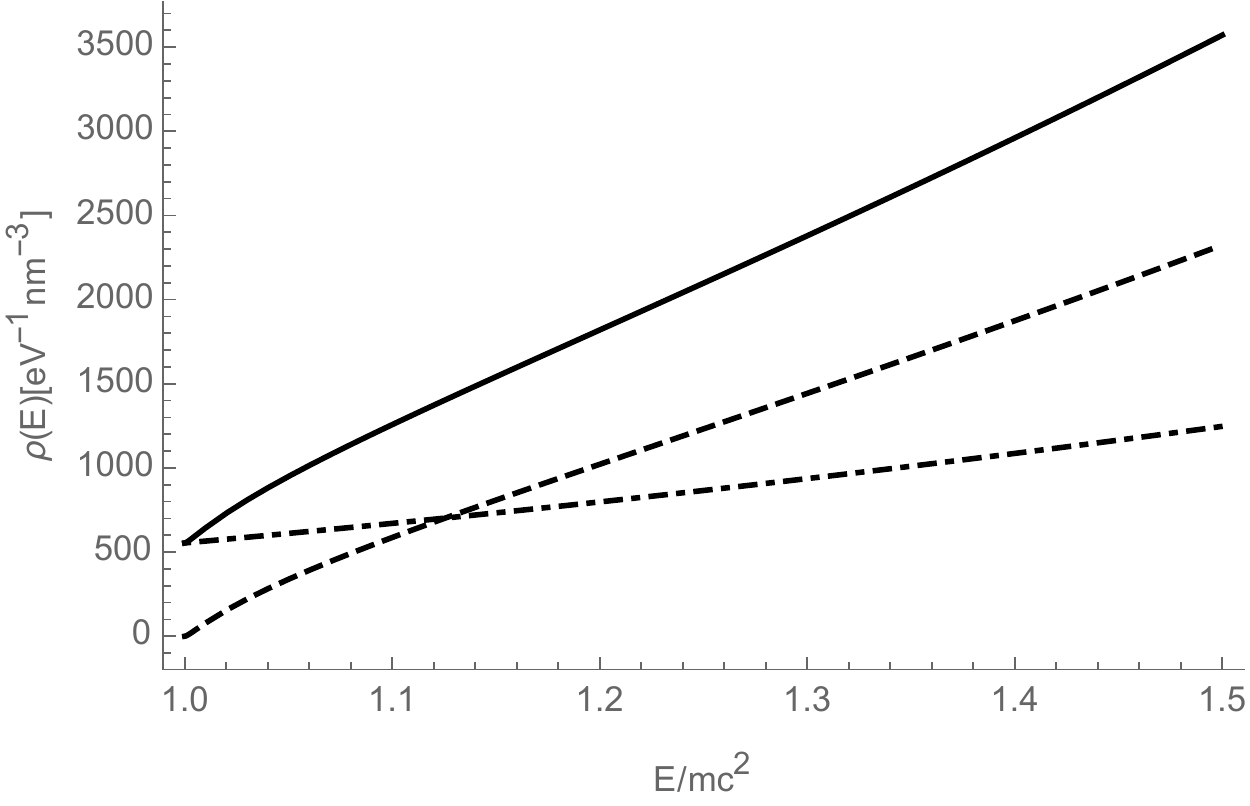}
		\label{fig:1}
		\caption{The density of states for charged bosonic particles of mass $m = m_e = .511 \text{ MeV}$ in the interface with $\mathcal{W} = 0.02 \, \mu\text{m} \cdot \text{eV}$. 
DotDashed: contribution from the bound states. 
Dashed: three-dimensional DOS in absence of any quantum wells.
Solid: the inter-dimensional DOS according to \eqref{KG DOS} }

\end{figure}

\begin{figure}[h]
		\centering
		\includegraphics[scale=.6]{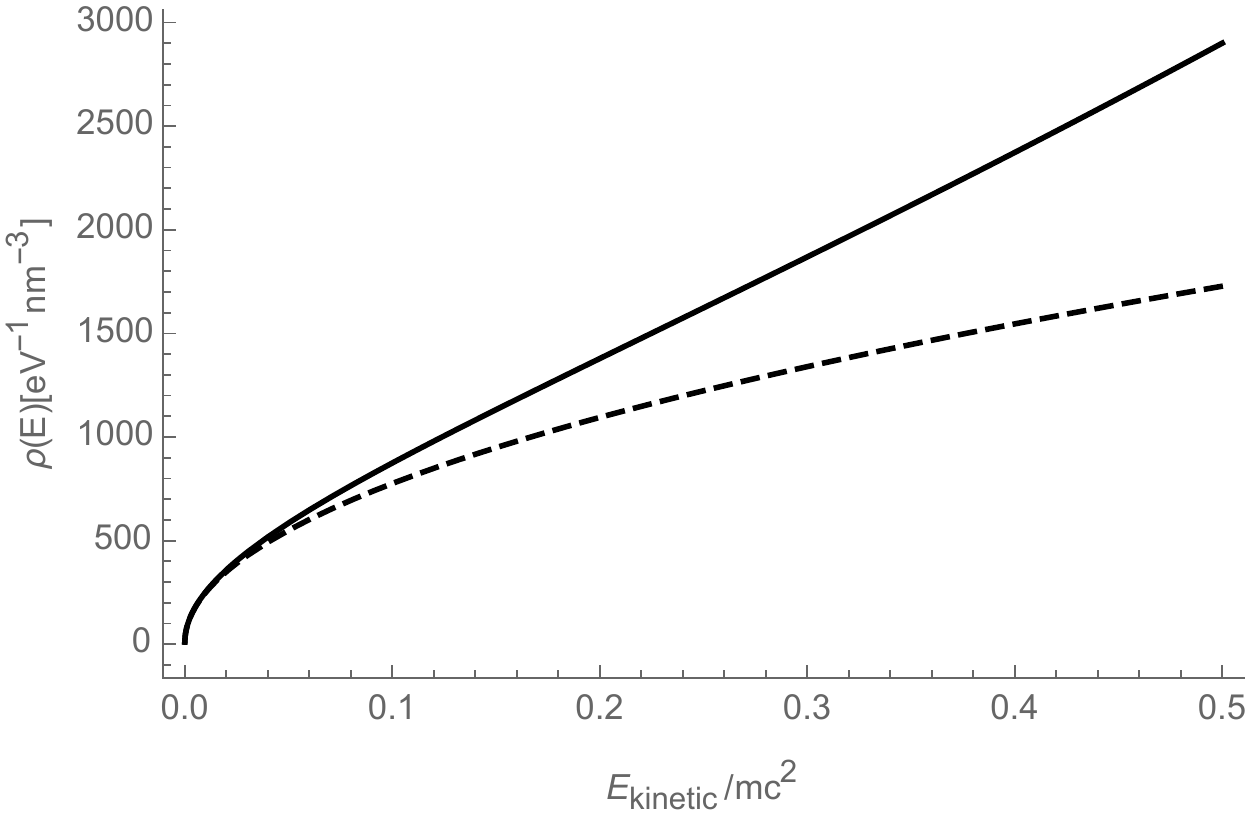}
		\label{fig:2}
		\caption{The density of states for charged bosonic particles of mass $m = m_e = .511 \text{ MeV}$ 
in the interface with $\mathcal{W} = 0.39 \, \text{nm} \cdot \text{eV}$ which corresponds to a binding energy of 1 eV. 
Dashed: the non-relativistic DOS given by equation \eqref{SE DOS}. 
Solid: the relativistic DOS given by \eqref{KG DOS}.}
\end{figure}

In the limit $\mathcal{W}\rightarrow 0$, the inter-dimensional DOS reduces to the three-dimensional relativistic DOS.
 For $\mathcal{W} > \hbar c$ the contribution from the two-dimensional term is enhanced so that in the limit $\mathcal{W} \gg \hbar c$, 
equation \eqref{KG DOS} tends to the two-dimensional term $\left(1+\frac{\mathcal{W}^{2}}{\hbar^{2}c^{2}}\right)^{-1}\frac{\mathcal{W}\tilde{E}}{\hbar^{2}c^{2}} \rho_{n=2}(\tilde{E})$
since the DOS is dominated by well states. 
In the low energy limit, $E\rightarrow mc^{2}$, equation \eqref{KG DOS} reduces to the non-relativistic equation from Schr\"{o}dinger 
theory \eqref{SE DOS}, Fig. 2. Furthermore, in the energy range for bound particle states
\begin{equation}
\frac{mc^2}{\sqrt{1+(\mathcal{W}/\hbar c)^2}}\le E<mc^2
\end{equation}
only the two-dimensional term in \eqref{KG DOS} contributes. All these limits behave as expected.

\section{\label{sec:level3}Shallow Finite Square Wells}

While delta wells are interesting, finite square wells provide a more realistic model for a low-dimensional systems as no real 
physical system is infinitely thin/genuinely two-dimensional. 

Although the Green's function method is useful for delta wells, when applied to finite square wells this method results in 
ambiguities in choosing how to shift the poles of the Green's function, thus motivating us to take a different approach. 

In general, the density of states is given by 
\begin{equation} \label{DOS formula}
\rho(E) dE = \frac{dn}{\mathcal{V}} = |\langle\boldsymbol{x}|\alpha\rangle|^2 d\alpha
\end{equation}
where $\alpha$ is the set of quantum numbers representing the state [1]. Therefore, to determine the DOS in systems with finite 
square wells, we construct the states of the system by solving the equation of motion for our model inter-dimensional system 
and then use equation \eqref{DOS formula}. 

We consider a system consisting of a three-dimensional bulk with an interface of width $2a$ centred at $z=0$. The electrostatic 
potential inside the interface differs by a constant, $\Phi_0$ (i.e., $A^\mu=(\Phi(\boldsymbol{x})/c,\boldsymbol{0})$ 
and $\Phi(\boldsymbol{x})=-\Phi_0\Theta(a^2-z^2)$). Substitution of $A^\mu$ into the equation of motion for coupling of charged 
Klein-Gordon and electromagnetic fields, equation \eqref{KG eq}, yields 
\begin{equation} \label{FW eq of motion}
\bigg( -\partial_0^2 + \nabla^2 - \frac{m^2c^2}{\hbar^2} + \Theta(a^2-z^2) \bigg( \frac{2i\mathcal{W}}{\hbar c}\partial_0 
+ \frac{\mathcal{W}^2}{\hbar^2c^2} \bigg) \bigg) \phi (\boldsymbol{x} ) = 0
\end{equation}
where $\mathcal{W} = q\Phi_0$. Similar to the delta well, $\mathcal{W}$, parametrizes the depth of the well. However, 
the factor $l$ is absent because the well thickness is now contained in the Heaviside step function, $\Theta(a^2-z^2)$.

Breaking up the spatial vector into components parallel to the interface, $\mathbf{x}_\parallel$, and perpendicular to 
the interface, $z$, we let $\mathbf{x} = (\mathbf{x}_\parallel,z)$. Then equation \eqref{FW eq of motion} can be solved 
by invoking several separation {\it ans\"atze} so that $\phi(x) = \phi_0(x^0) \phi_\parallel (\mathbf{x}_\parallel) \phi_\perp (z)$. 
The parallel and time components of the wave function are easily found to be $\phi_\parallel (\mathbf{x}_\parallel) 
= \frac{1}{2 \pi} e^{i \mathbf{k}_\parallel \cdot \mathbf{x}_\parallel}$ and $\phi_0(x^0) = e^{-i k^0 x^0}$. 

Note that $\phi_\parallel$ is a normalized plane wave in the plane parallel to the interface. Also,  $\phi_0$ does not 
contain a factor of $1/\sqrt{2 \pi}$ because wave functions cannot be normalized in the time direction.

The differential equation for the perpendicular wave function is given by
\begin{eqnarray} \label{perp ode}
\frac{\partial_\perp^2 \phi_\perp}{\phi_\perp} &=& - (k^0)^2 - \Theta(a^2-z^2) \frac{2\mathcal{W}}{\hbar c} k^0 - \Theta(a^2-z^2) \frac{\mathcal{W}^2}{\hbar^2c^2}  
\nonumber \\
&& + \frac{m^2 c^2}{\hbar^2} + k_\parallel^2 = \pm k_\perp^2.
\end{eqnarray}
Inside the well it has the possible solutions
\begin{eqnarray}
\phi_\perp(z) &=& A \sin(k_\perp z) + B \cos(k_\perp z)
\\
\phi_\perp(z) &=& A e^{k_\perp z} + B e^{-k_\perp z}
\end{eqnarray}
with energies
\begin{eqnarray}
E &=& - \mathcal{W} \pm \, \sqrt{m^2c^4 + \hbar^2 c^2 k_\parallel^2 \pm \hbar^2 c^2 k_\perp^2};
\end{eqnarray}
outside the well, the solutions are
\begin{eqnarray}
\phi_\perp(z) &=& A \sin(k^\prime_\perp z) + B \cos(k^\prime_\perp z)
\\
\phi_\perp(z) &=& A e^{k^\prime_\perp z} + B e^{-k^\prime_\perp z}
\end{eqnarray}
with energies
\begin{eqnarray}
E &=& \pm \, \sqrt{m^2c^4 + \hbar^2 c^2 k_\parallel^2 \pm \hbar^2 c^2 {k^\prime_\perp}^2}.
\end{eqnarray}
Here, $k_\perp$ denotes the perpendicular component of momentum inside the well while $k_\perp^\prime$ denotes the perpendicular 
component of momentum outside the well.

We consider the simplest case of shallow wells, $0 \leq \mathcal{W} \leq mc^2$, which yield free particle and anti-particle, 
bound particle, and tunneling anti-particle states. These wave functions were calculated and are listed in appendix B. 

There is no mixing of particle and anti-particle states (this requires $\mathcal{W} > mc^2$) nor particle anti-particle pair 
production at the well boundaries (i.e. Klein paradox which requires $\mathcal{W} > 2mc^2$). 
Additionally, charge is conserved for Klein-Gordon particles.
Therefore, we can form a well defined DOS even though the Klein-Gordon field does not have a conserved probability density. 
The DOS for shallow wells (given by equation \eqref{DOS formula}) is then found to be
\begin{multline} \label{DOS final}
\rho(E, \boldsymbol{x}_\parallel, z) =  \rho(E, z) = 
 \\
 \sum_\pm \sum_n \Theta \Big(E + \mathcal{W} - \sqrt{m^2 c^4 +\hbar^2 c^2 \zeta^2_n/a^2}\Big)  \frac{E + \mathcal{W}}{2 \pi \hbar^2 c^2} |\psi^\pm_n(z)|^2
 \\
+ \sum_\pm  \Theta \Big(E- mc^2\Big) \frac{E + \mathcal{W}}{2 \pi \hbar^2 c^2} \int^{*} dk_\perp |\psi^\pm_{\text{part}}(z)|^2
 \\
+ \sum_\pm  \Theta \Big(-mc^2 - (E+\mathcal{W}) \Big) \frac{E + \mathcal{W}}{2 \pi  \hbar^2 c^2} \int^{*}dk_\perp  |\psi^\pm_{\text{anti}}(z)|^2
 \\
+ \sum_\pm \Theta\Big(-mc^2-E\Big)  \frac{E + \mathcal{W}}{2 \pi  \hbar^2 c^2} \int^{*} d\kappa_\perp |\psi^\pm_{\text{tunnel}}(z)|^2
\end{multline}
where $(+)$ denotes the positive parity states and $(-)$ the negative parity states. The integration limits, the dependences of
the wave functions on the wavenumbers, and the bound state parameters $\zeta_n$ are given in Appendix B. 

Equation \eqref{DOS final} contains contributions from both particle and anti-particle states. To obtain the DOS for either particles or 
anti-particles, we simply omit the undesired states from equation \eqref{DOS final}. The integrals in \eqref{DOS final} cannot be solved analytically 
and therefore were computed numerically.

\begin{figure}[h]
\centering
\includegraphics[scale=.6]{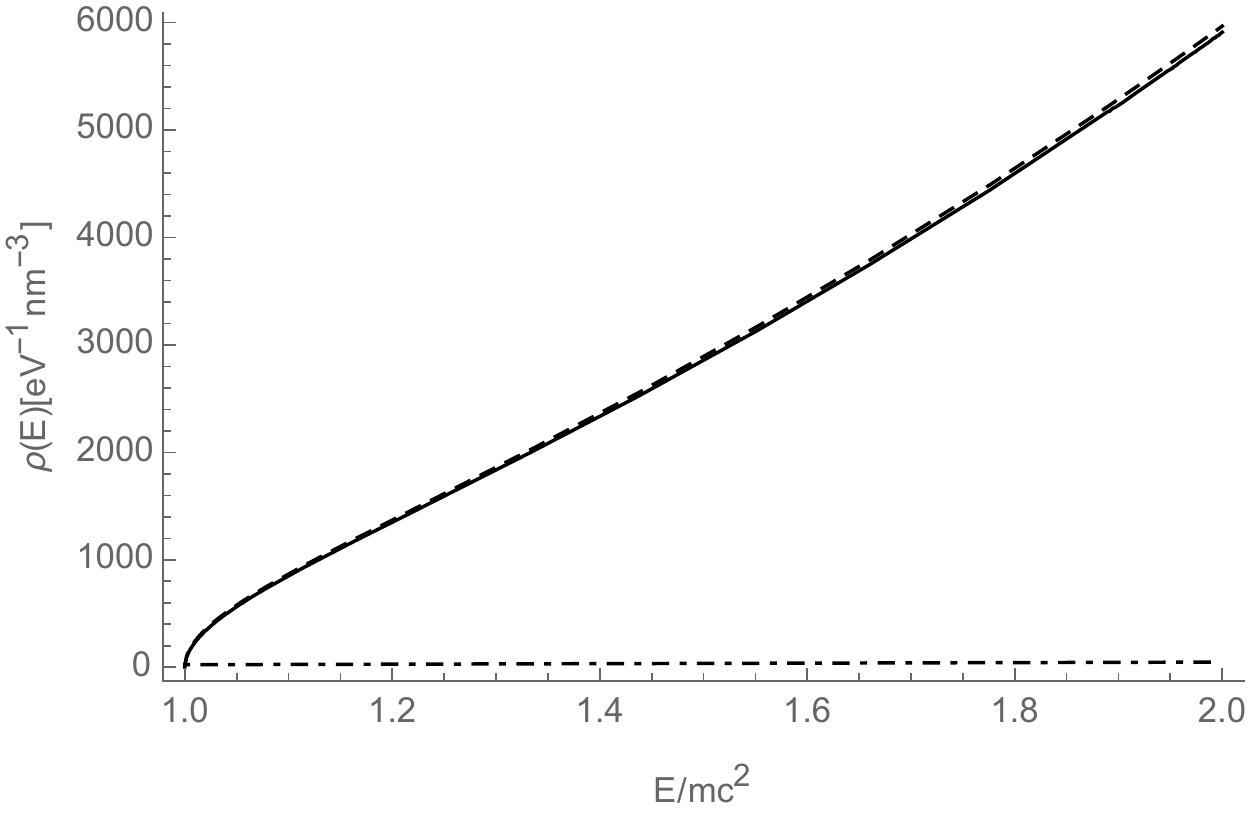}
\caption{The particle DOS in the interface at $z=0$ for $\mathcal{W} = mc^2/10000$ and mass $m = m_e = 0.511 \text{MeV}$. 
The width was set to 1 micrometer.
DotDashed: is the contribution from the states bound inside the quantum well. 
Dashed: is the three-dimensional DOS in absence of any quantum wells. 
Solid: is the DOS according to \eqref{DOS final}.}
\end{figure}

\begin{figure}[h]
\centering
\includegraphics[scale=.6]{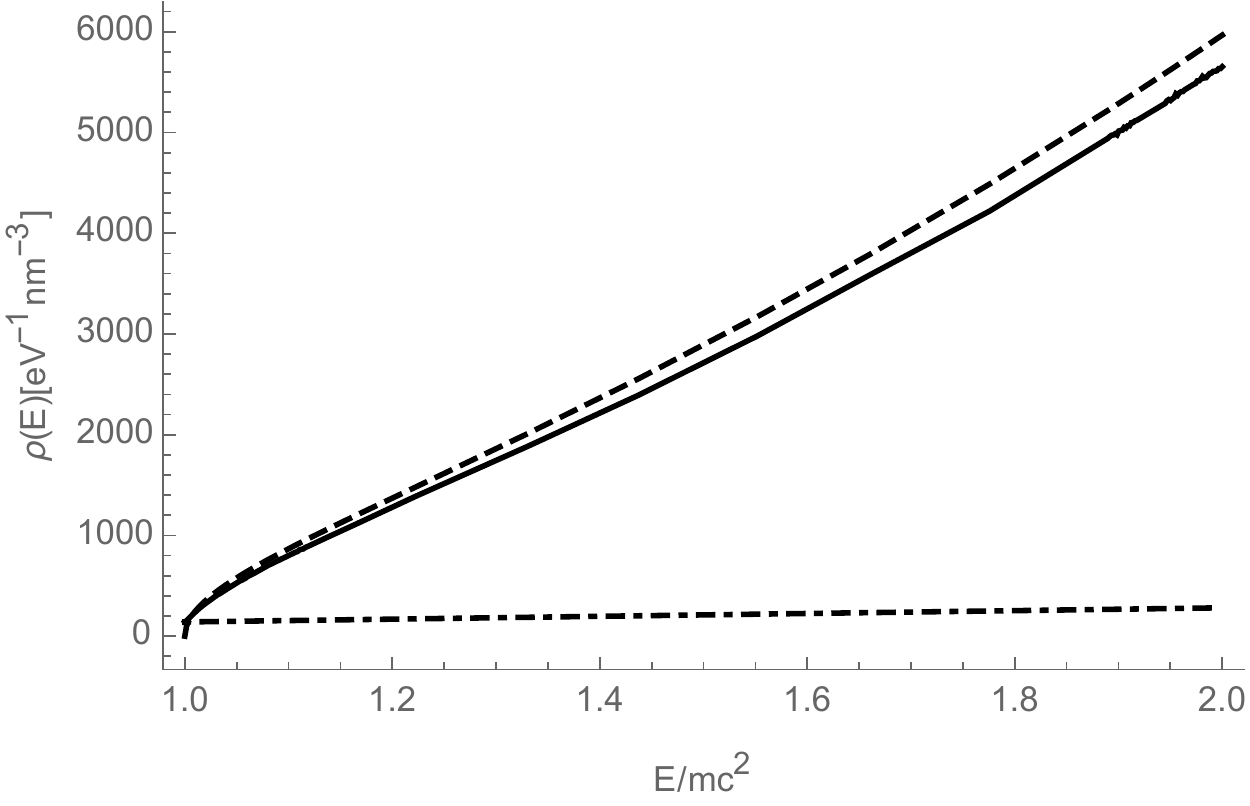}
\caption{The particle DOS in the interface at $z=0$ for $\mathcal{W} = mc^2/300$ and mass $m = m_e = 0.511 \text{MeV}$. 
The width was set to 1 micrometer.
DotDashed: is the contribution from the states bound inside the quantum well. 
Dashed: is the three-dimensional DOS in absence of any quantum wells. 
Solid: is the DOS according to \eqref{DOS final}.}
\end{figure}

The DOS in inter-dimensional systems with shallow wells is an interpolation between the well known two- and three-dimensional DOS.
 For extremely shallow well depths, $\mathcal{W} \ll mc^2$, the inter-dimensional DOS approaches the three-dimensional limit, Fig. 3; 
increasing the well depth brings the inter-dimensional DOS closer toward the two-dimensional DOS, Fig. 4. 
This is a consequence of larger number of states in the well, which corresponds to an increase of the relative weight
of the two-dimensional contribution from the well states to the DOS.

Using a similar procedure as outlined in this section we also calculated the inter-dimensional DOS for finite square wells 
in Schr\"{o}dinger theory. In the low-energy limit, $E \rightarrow mc^2$, we observe agreement between the Klein-Gordon and 
Schr\"{o}dinger inter-dimensional DOS.

\section{\label{sec:level4}Conclusions}
We have calculated the density of states for quasi-relativistic bosons moving in systems with thin interfaces, where we have assumed 
that the electrostatic potential differs by a constant from the bulk material. We have found that the density of states in the well 
reduces to the non-relativistic density of states from Schr\"{o}dinger theory in the low energy limit, $E \rightarrow mc^2$. Also in 
accord with intuition, the inter-dimensional density of states tends to the three-dimensional density of states in the 
limit $\mathcal{W}\rightarrow 0$, whereas the contribution from the two-dimensional density of well states 
increases for larger $\mathcal{W}$, corresponding in particular to larger well depth. Furthermore, we found that the dimensional
conversion factor $\kappa$ (inverse penetration length of the well states) which converts the two-dimensional density of states
in the energy scale to a contribution to the three-dimensional density of states, in the relativistic case 
becomes $\kappa/\sqrt{1+(\mathcal{W}/\hbar c)^2}$.


\section*{ACKNOWLEDGMENTS}
This work was supported in part by NSERC Canada.

\section*{\label{sec:A}Appendix A: Solution to Equation (6)}
Substitution of the Fourier decomposition of $\left\langle x|G|x^{\prime}\right\rangle $,
\begin{multline}
\left\langle x|G|x^{\prime}\right\rangle =\int d^{2}\boldsymbol{k}_{\parallel}\int d^{2}\boldsymbol{k^{\prime}}_{\parallel}\int dk^{0}\int dk^{\prime0}\int dk_{\perp}
 \\
\times  \exp\left[i\left(k^{\prime 0}x^{\prime 0}-k^{0}x^{0}+\boldsymbol{k}_{\parallel}\cdot\boldsymbol{x}_{\parallel}
-\boldsymbol{k}_{\parallel}^{\prime}\cdot\boldsymbol{x}_{\parallel}^{\prime}+k_{\perp}z\right)\right]
\\
\times  \left(\frac{1}{2\pi}\right)^{\frac{7}{2}} \left\langle k^{0},\boldsymbol{k}_{\parallel},k_{\perp}|G|k^{\prime0},
\boldsymbol{k}_{\parallel}^{\prime},z^{\prime}\right\rangle,
\end{multline}
into equation \eqref{KG Greens} yields 
\begin{multline}
\left(\frac{1}{2\pi}\right)^{\frac{7}{2}}\int d^{2}\boldsymbol{k}_{\parallel}\int d^{2}\boldsymbol{k}_{\parallel}^{\prime}
\int dk^{0}\int dk^{\prime0}\int dk_{\perp} 
\\
\times \left\{ (k^{0})^{2}-k_{\parallel}^{2}-k_{\perp}^{2}-\frac{m^{2}c^{2}}{\hbar^{2}}+2\frac{ql\Phi_0}{c\hbar}k^{0}\delta(z-z_{0})\right\} 
 \\
\times\exp\left[i\left(k^{\prime 0}x^{\prime 0}-k^{0}x^{0}+\boldsymbol{k}_{\parallel}\cdot\boldsymbol{x}_{\parallel}-\boldsymbol{k}_{\parallel}^{\prime}
\cdot\boldsymbol{x}_{\parallel}^{\prime}+k_{\perp}z\right)\right] 
\\
\times \left\langle k^{0},\boldsymbol{k}_{\parallel},k_{\perp}|G|k^{\prime0},\boldsymbol{k}_{\parallel}^{\prime},z^{\prime}\right\rangle 
 \\
=-\delta(z-z^{\prime})\delta(x^{0}-x^{\prime0})\delta(\boldsymbol{x}_{\parallel}-\boldsymbol{x}_{\parallel}^{\prime}).
\end{multline}
Next, we take the inverse Fourier transform with respect to the variables
$x^{0},x^{\prime0},\boldsymbol{x}_{\parallel},\boldsymbol{x}_{\parallel}^{\prime}$
and evaluate the resulting integrals to get
\begin{multline}
\int dk_{\perp}\exp\left(ik_{\perp}z\right) \times\left\langle \kappa^{0},\boldsymbol{\kappa}_{\parallel},k_{\perp}|G|\kappa^{\prime0},
\boldsymbol{\kappa}_{\parallel}^{\prime},z^{\prime}\right\rangle
\\
\left\{ \frac{m^{2}c^{2}}{\hbar^{2}}+\kappa_{\parallel}^{2}+k_{\perp}^{2}-(\kappa^{0})^{2}-2\frac{ql\Phi_0}{c\hbar}\kappa^{0}\delta(z-z_{0})\right\} 
\\
 =\sqrt{2\pi}\delta(z-z^{\prime})\delta(\boldsymbol{k}_{\parallel}-\boldsymbol{k}_{\parallel}^{\prime})\delta\left(k^{0}-k^{\prime0}\right).
\end{multline}
Taking the inverse Fourier Transform with respect to $z$, we get
\begin{multline} \label{invFT wrt z}
\int dk_{\perp}\int dz\exp\left[iz\left(k_{\perp}-\kappa_{\perp}\right)\right]\left\langle 
\kappa^{0},\boldsymbol{\kappa}_{\parallel},k_{\perp}|G|\kappa^{\prime0},\boldsymbol{\kappa}_{\parallel}^{\prime},z^{\prime}\right\rangle \\
\times\left\{ \frac{m^{2}c^{2}}{\hbar^{2}}+\kappa_{\parallel}^{2}+k_{\perp}^{2}-(\kappa^{0})^{2}-2\frac{ql\Phi_0}{c\hbar}\kappa^{0}\delta(z-z_{0})\right\} \\
=\sqrt{2\pi}\int dz\exp\left(-i\kappa_{\perp}z\right)\delta(z-z^{\prime})\delta(\boldsymbol{k}_{\parallel}-\boldsymbol{k}_{\parallel}^{\prime})\delta\left(k^{0}-k^{\prime0}\right).
\end{multline}
Since \eqref{KG Greens} is a linear differential equation of constant coefficients with respect to all variables with the exception of $z\text{ and }z^{\prime}$, it follows
that $\left\langle k^{0},\boldsymbol{k}_{\parallel},k_{\perp}|G|k^{\prime0},\boldsymbol{k}_{\parallel}^{\prime},z^{\prime}\right\rangle 
=\left\langle k_{\perp}|G(k^{0},\boldsymbol{k}_{\parallel})|z^{\prime}\right\rangle \delta(\boldsymbol{k}_{\parallel}-\boldsymbol{k}_{\parallel}^{\prime})
\delta\left(k^{0}-k^{\prime0}\right)$. Equation \eqref{invFT wrt z}  becomes
\begin{multline} \label{full FT mixed}
\left( \frac{m^{2}c^{2}}{\hbar^{2}}+\kappa_{\parallel}^{2}+\kappa_{\perp}^{2}-(\kappa^{0})^{2}\right) 
\left\langle \kappa_{\perp}|G(\kappa^{0},\boldsymbol{\kappa}_{\parallel})|z^{\prime}\right\rangle \exp\left(iz_{0}\kappa_{\perp}\right)
\\
-\frac{ql\Phi_0}{c\hbar}\kappa^{0}\int\frac{dk_{\perp}}{\pi}\exp\left(iz_{0}k_{\perp}\right)\left\langle k_{\perp}|G(\kappa^{0},
\boldsymbol{\kappa}_{\parallel})|z^{\prime}\right\rangle 
\\
=\frac{1}{\sqrt{2\pi}}\exp\left[i\kappa_{\perp}\left(z_{0}-z^{\prime}\right)\right].
\end{multline}
The mixed representation, $\left\langle \kappa_{\perp}|G(\kappa^{0},\boldsymbol{\kappa}_{\parallel})|z^{\prime}\right\rangle ,$
appears both inside and outside of the integral in equation \eqref{full FT mixed}. Since
we are integrating over the perpendicular momentum variable, $\int\frac{dk_{\perp}}{\pi}\exp\left(iz_{0}k_{\perp}\right)
\left\langle k_{\perp}|G(\kappa^{0},\boldsymbol{\kappa}_{\parallel})|z^{\prime}\right\rangle $
is a function of $\kappa^{0},\boldsymbol{\kappa}_{\parallel},\text{ and }z^{\prime}$.
This suggests writing \eqref{full FT mixed} as
\begin{multline}
\left\langle \kappa_{\perp}|G(\kappa^{0},\boldsymbol{\kappa}_{\parallel})|z^{\prime}\right\rangle \exp\left(iz_{0}\kappa_{\perp}\right)=
\\
\frac{\frac{\exp\left[i\kappa_{\perp}\left(z_{0}-z^{\prime}\right)\right]}{\sqrt{2\pi}}+f(\kappa^{0},
\boldsymbol{\kappa}_{\parallel},z^{\prime})}{\frac{m^{2}c^{2}}{\hbar^{2}}+\kappa_{\parallel}^{2}+\kappa_{\perp}^{2}-(\kappa^{0})^{2}-i\epsilon},
\end{multline}
where we have shifted the poles in agreement with the conventions for a retarded Green's function
and $f(\kappa^{0},\boldsymbol{\kappa}_{\parallel},z^{\prime})$
satisfies
\begin{multline}
f(\kappa^{0},\boldsymbol{\kappa}_{\parallel},z^{\prime})+\frac{ql\Phi_0}{c\hbar}\kappa^{0}\int\frac{dk_{\perp}}{\pi}\exp\left(iz_{0}k_{\perp}\right)
\\
\times \left\langle k_{\perp}|G(\kappa^{0},\boldsymbol{\kappa}_{\parallel})|z^{\prime}\right\rangle =0.
\end{multline}
Solving for $f(k^{0},\boldsymbol{k}_{\parallel,}z^{\prime})$, we
find
\begin{equation} \label{def f}
f(k^{0},\boldsymbol{k}_{\parallel},z^{\prime})=\frac{\frac{1}{\sqrt{2\pi}}\frac{ql\Phi_0}{c\hbar}k^{0}\int\frac{d\kappa_{\perp}}{\pi}
\frac{\exp\left[i\kappa_{\perp}\left(z_{0}-z^{\prime}\right)\right]}{\kappa_{\perp}^{2}-\left((k^{0})^{2}-\frac{m^{2}c^{2}}{\hbar^{2}}-k_{\parallel}^{2}\right)
-i\epsilon}}{1-\frac{ql\Phi_0}{c\hbar}k^{0}\int\frac{d\kappa_{\perp}}{\pi}\frac{1}{\kappa_{\perp}^{2}-\left((k^{0})^{2}
-\frac{m^{2}c^{2}}{\hbar^{2}}-k\parallel\right)-i\epsilon}}.
\end{equation}
Simplifying the integrals in \eqref{def f} yields
\begin{multline}
f(k^{0},\boldsymbol{k}_{\parallel},z^{\prime})=\frac{\Theta\left((k^{0})^{2}-k_{\parallel}^{2}-\frac{m^{2}c^{2}}{\hbar^{2}}\right)}{\sqrt{2\pi}} 
\\
\times \left(\frac{\frac{ql\Phi_0}{c\hbar}\kappa^{0}i\exp\left[i\left((k^{0})^{2}-k_{\parallel}^{2}-\frac{m^{2}c^{2}}{\hbar^{2}}\right)
\left|z_{0}-z^{\prime}\right|\right]}{\sqrt{(k^{0})^{2}-k_{\parallel}^{2}-\frac{m^{2}c^{2}}{\hbar^{2}}}-i\frac{ql\Phi_0}{c\hbar}\kappa^{0}}\right)\\
+\frac{\Theta\left(\frac{m^{2}c^{2}}{\hbar^{2}}+k_{\parallel}^{2}-(k^{0})^{2}\right)}{\sqrt{2\pi}}
\\
\times \left(\frac{\frac{ql\Phi_0}{c\hbar}\kappa^{0}\exp\left[-\left(\frac{m^{2}c^{2}}{\hbar^{2}}+k_{\parallel}^{2}-(k^{0})^{2}\right)
\left|z_{0}-z^{\prime}\right|\right]}{\sqrt{\frac{m^{2}c^{2}}{\hbar^{2}}+k_{\parallel}^{2}-(k^{0})^{2}}-\frac{ql\Phi_0}{c\hbar}\kappa^{0}-i\epsilon}\right).
\end{multline}

Hence, we find that 
\begin{multline} 
\left\langle k_{\perp}|G(k^{0},\boldsymbol{k}_{\parallel})|z^{\prime}\right\rangle = \frac{1}{\sqrt{2\pi}} \frac{1}{\frac{m^{2}c^{2}}{
\hbar^{2}}+k_{\perp}+k_{\parallel}^{2}-(k^{0})^{2}-i\epsilon}
\\
\times \left\{ \exp\left(-ikz^{\prime}\right)+\Theta\left((k^{0})^{2}-k_{\parallel}^{2}-\frac{m^{2}c^{2}}{\hbar^{2}}\right)\exp\left(-ik_{\perp}z_{0}\right)\right.\\
\times\left(\frac{\frac{ql\Phi_0}{c\hbar}k^{0}i\exp\left[i\left((k^{0})^{2}-k_{\parallel}^{2}-\frac{m^{2}c^{2}}{\hbar^{2}}\right)\left|z_{0}-z^{\prime}\right|\right]
}{\sqrt{(k^{0})^{2}-k_{\parallel}^{2}-\frac{m^{2}c^{2}}{\hbar^{2}}}-i\frac{ql\Phi_0}{c\hbar}k^{0}}\right)\\
+\Theta\left(\frac{m^{2}c^{2}}{\hbar^{2}}+k_{\parallel}^{2}-(k^{0})^{2}\right)\exp\left(-ik_{\perp}z_{0}\right)\\
\left.\times\left(\frac{\frac{ql\Phi_0}{c\hbar}k^{0}\exp\left[-\left(\frac{m^{2}c^{2}}{\hbar^{2}}+k_{\parallel}^{2}-(k^{0})^{2}\right)\left|z_{0}-z^{\prime}\right|
\right]}{\sqrt{\frac{m^{2}c^{2}}{\hbar^{2}}+k_{\parallel}^{2}-(k^{0})^{2}}-\frac{ql\Phi_0}{c\hbar}k^{0}-i\epsilon}\right)\right\} .
\end{multline}

To obtain $\left\langle z|G(E,\boldsymbol{x}_{\parallel})|z^{\prime}\right\rangle |_{E=\hbar ck^{0}}$,
we first preform an inverse Fourier transform with respect to $z^{\prime}$
on the mixed representation of the retarded Green's function, $\left\langle k_{\perp}|G(k^{0},
\boldsymbol{k}_{\parallel})|z^{\prime}\right\rangle $,
to get
\begin{multline}
\left\langle z|G(E,\boldsymbol{k}_{\parallel})|z^{\prime}\right\rangle |_{E=\hbar ck^{0}} = \frac{\Theta\left((k^{0})^{2}-k_{\parallel}^{2}-\frac{m^{2}c^{2}}{\hbar^{2}}
\right)}{2\left(\sqrt{(k^{0})^{2}-k_{\parallel}^{2}-\frac{m^{2}c^{2}}{\hbar^{2}}}\right)} 
\\
\times i \left\{ \exp\left[i\sqrt{(k^{0})^{2}-k_{\parallel}^{2}-\frac{m^{2}c^{2}}{\hbar^{2}}}\left|z-z^{\prime}\right|\right] \right. +\frac{ql\Phi_0}{c\hbar}k^{0}
 \\
\times \left. i\frac{\exp\left[i\sqrt{(k^{0})^{2}-k_{\parallel}^{2}-\frac{m^{2}c^{2}}{\hbar^{2}}}\left(\left|z-z_{0}\right|+\left|z^{\prime}-z_{0}\right|\right)
\right]}{\sqrt{(k^{0})^{2}-k_{\parallel}^{2}-\frac{m^{2}c^{2}}{\hbar^{2}}}-i\frac{ql\Phi_0}{c\hbar}\kappa^{0}}\right\} 
 \\
+ \frac{\Theta\left(\frac{m^{2}c^{2}}{\hbar^{2}}+k_{\parallel}^{2}-(k^{0})^{2}\right)}{2\left(\sqrt{\frac{m^{2}c^{2}}{\hbar^{2}}+k_{\parallel}^{2}-(k^{0})^{2}}\right)}
\\
\times \left\{ \exp\left[-\sqrt{\frac{m^{2}c^{2}}{\hbar^{2}}+k_{\parallel}^{2}-(k^{0})^{2}}\left|z-z^{\prime}\right|\right]\right. + \frac{ql\Phi_0}{c\hbar}k^{0}
 \\
\times \left.\frac{\exp\left[-\sqrt{\frac{m^{2}c^{2}}{\hbar^{2}}+k_{\parallel}^{2}-(k^{0})^{2}}\left(\left|z-z_{0}\right|+\left|z^{\prime}-z_{0}\right|\right)
\right]}{\sqrt{\frac{m^{2}c^{2}}{\hbar^{2}}+k_{\parallel}^{2}-(k^{0})^{2}}-\frac{ql\Phi_0}{c\hbar}\kappa^{0}-i\epsilon}\right\} .
\end{multline}
Performing a further Fourier transform with respect to $\boldsymbol{k}_{\parallel}$
yields 
\begin{multline}
\left\langle \boldsymbol{x}|G(E)|\boldsymbol{x}^{\prime}\right\rangle = \int d^{2}\boldsymbol{k}_{\parallel}\exp\left[i\boldsymbol{k}_{\parallel}
\left(\boldsymbol{x}_{\parallel}-\boldsymbol{x}_{\parallel}^{\prime}\right)\right]
\\
\times \left\langle z|G(E,\boldsymbol{k}_{\parallel})|z^{\prime}\right\rangle
 \\
=\int d\theta\int\frac{dk_{\parallel}}{(2\pi)^{2}}k_{\parallel}\exp\left[ik_{\parallel}\left|x_{\parallel}-x_{\parallel}^{\prime}\right|\cos(\theta)\right] 
\\
\times \left\langle z|G(E,\boldsymbol{k}_{\parallel})|z^{\prime}\right\rangle 
\\
=\int\frac{dk_{\parallel}}{2\pi}J_{0}(k_{\parallel}\left|x_{\parallel}-x_{\parallel}^{\prime}\right|)k_{\parallel}\left\langle z|G(E,\boldsymbol{k}_{\parallel})|z^{\prime}\right\rangle .
\end{multline}
However, we are most interested in the Green's function at the interface, $z = z^\prime$, where we expect the most prominent inter-dimensional effects. 
At the interface, $J_{0}(0)=1$, and 
\begin{equation}
\left\langle \boldsymbol{x}|G(E)|\boldsymbol{x}\right\rangle |_{E=\hbar ck^{0}}=\int\frac{dk_{\parallel}}{2\pi}k_{\parallel}\left\langle z|G(E,\boldsymbol{k}_{\parallel})|z\right\rangle .
\end{equation}

\section*{\label{sec:B}Appendix B: Summary of States }
In this appendix we collect the wave functions with their appropriate energy and momentum ranges, necessary to carry out the integration in equation \eqref{DOS final}.
\subsection*{Free Particle and Anti-Particle States}
The wave functions for the free particle and anti-particle states are
\begin{multline}
\phi_{part}^+(z) = \phi_{anti}^+ (z) = \frac{1 / \sqrt{\pi}}{  \sqrt{ \cos^2(k_\perp a) + \frac{k_\perp^2}{{k_\perp^\prime}^2} \sin^2(k_\perp a) }}  
\\
\times \bigg[ \Theta(a^2-z^2) \cos(k_\perp z) + \Theta(z^2-a^2)
 \\
 \times \Big[  \cos(k_\perp a) \cos\left(k_\perp^\prime (|z|-a)\right) 
 \\
 - \frac{k_\perp}{k_\perp^\prime} \sin(k_\perp a) \sin\left(k_\perp^\prime (|z|-a)\right) \Big] \bigg]
\end{multline}
and
\begin{multline}
\phi_{part}^-( z) = \phi_{anti}^- (z) = \frac{1/ \sqrt{\pi}}{\sqrt{ \sin^2(k_\perp a) + \frac{k_\perp^2}{{k_\perp^\prime}^2} \cos^2(k_\perp a) }}  
\\
\times \bigg[ \Theta(a^2-z^2) \sin(k_\perp z)  + \Theta(z^2-a^2)  
\\
\times \text{sign}(z) \Big[ \sin(k_\perp a) \cos\left( k_\perp^\prime (|z|-a) \right) 
\\
+ \cos(k_\perp a) \sin\left(k_\perp^\prime (|z|-a) \right) \Big] \bigg].
\end{multline}
The free particle energy is given by
\begin{eqnarray}
E &=& -\mathcal{W} + \sqrt{m^2 c^4 + \hbar^2 c^2 (k^2_\parallel + k^2_\perp)} 
\nonumber \\
&=& \sqrt{m^2 c^4 + \hbar^2 c^2 (k^2_\parallel + {k^\prime_\perp}^2)} .
\end{eqnarray}
where $E\geq mc^2$. Therefore, for fixed energy, $k_\perp$ is restricted to the interval 
\begin{equation}
\sqrt{2E\mathcal{W}+\mathcal{W}^2 } \leq \hbar c k_\perp \leq \sqrt{(E+\mathcal{W})^2-m^2c^4}.
\end{equation}

The free anti-particle energy is given by
\begin{eqnarray}
\bar{E} &=& \mathcal{W} + \sqrt{m^2 c^4 + \hbar^2 c^2 (k^2_\parallel + k^2_\perp)} 
\nonumber \\
&=& \sqrt{m^2 c^4 + \hbar^2 c^2 (k^2_\parallel + {k^\prime_\perp}^2)} .
\end{eqnarray}
where $\bar{E}\geq mc^2+\mathcal{W}$. For fixed energy, $k_\perp$ is restricted to the interval 
\begin{equation}
0 \leq \hbar c k_\perp \leq \sqrt{(E+\mathcal{W})^2-m^2c^4}.
\end{equation}

\subsection*{Tunneling Anti-Particle States}
The wave functions for the tunneling anti-particle states are
\begin{multline}
\phi_{tunnel}^+(z) = \frac{1 / \sqrt{\pi}}{\sqrt{ \cosh^2(k_\perp a) + \frac{k_\perp ^2}{{k_\perp ^\prime}^2} \sinh^2(k_\perp a)}} 
\\
\times \bigg[ \Theta(a^2-z^2) \cosh(k_\perp z) + \Theta(z^2-a^2) 
 \\
\times \Big[ \cosh (k_\perp a) \cos\left(k_\perp^\prime(|z|-a) \right) 
\\
+ \frac{k_\perp}{k_\perp^\prime} \sinh (k_\perp a) \sin\left( k^\prime_\perp (|z|-a)\right) \Big] \bigg] 
\end{multline}
and
\begin{multline}
\phi_{tunnel}^-(z) = \frac{1 / \sqrt{\pi}  }{\sqrt{\sinh^2(k_\perp a) + \frac{k_\perp^2}{{k_\perp^\prime}^2} \cosh^2(k_\perp a)}} 
\\
 \times \bigg[ \Theta(a^2-z^2) \sinh(k_\perp z) + \Theta(z^2-a^2)   
 \\
\times \text{sign}(z) \Big[ \sinh(k_\perp a) \cos\left( k_\perp^\prime (|z|-a) \right) 
\\
+ \frac{k_\perp}{k_\perp^\prime} \cosh(k_\perp a) \sin\left( k_\perp (|z|-a) \right) \Big] \bigg].
\end{multline}
The tunneling anti-particle energy is given by
\begin{eqnarray}
\bar{E} &=& \mathcal{W} - \sqrt{m^2 c^4 + \hbar^2 c^2 (k^2_\parallel - k^2_\perp)} 
\nonumber \\
&=& \sqrt{m^2 c^4 + \hbar^2 c^2 (k^2_\parallel + {k^\prime_\perp}^2)} 
\end{eqnarray}
where $\bar{E} \geq mc^2$ and for a fixed energy, $k_\perp$ is restricted to the interval 
\begin{equation}
\Re\sqrt{m^2c^4 - (E+\mathcal{W})^2} \leq \hbar c k_\perp \leq \sqrt{-(2E\mathcal{W} +\mathcal{W}^2)}.
\end{equation}

\subsection*{Bound Particle States} 
The wave functions for the bound particle states are
\begin{multline}
\phi_n^+(z) = \frac{ e^{\zeta^\prime_n} }{\sqrt{a}} \bigg[ 1 + \frac{1}{\zeta^\prime_n} \bigg]^{-1/2} 
 \bigg[ \Theta(a^2-z^2) \cos(\zeta_n z/a) e^{-\zeta^\prime_n} 
\\
+ \Theta(z^2-a^2) \cos(\zeta_n) e^{-\zeta^\prime_n |z|/a} \bigg]
\end{multline}
and 
\begin{multline}
\phi^-_n(z) = \frac{ e^{\zeta^\prime_n} }{\sqrt{a}} \bigg[ 1 + \frac{1}{\zeta^\prime_n} \bigg]^{-1/2} \bigg[ \Theta(a^2-z^2) \sin(\zeta_n z/a) e^{-\zeta^\prime_n} 
\\
+ \Theta(z^2-a^2) \text{sign}(z) \sin(\zeta_n) e^{-\zeta^\prime_n |z|/a} \bigg].
\end{multline}
For the even solution, the momentum perpendicular to the well is quantized by the condition 
\begin{multline}
\zeta_n \tan(\zeta_n) = \sqrt{\frac{2\mathcal{W}a}{\hbar c} \sqrt{\zeta^2_n+\frac{m^2 c^2}{\hbar^2} a^2} - \zeta^2_n - \frac{\mathcal{W}^2 a^2}{\hbar^2 c^2}} 
= \zeta^\prime_n
\end{multline}
where $\zeta = k_\perp /a$ and $\zeta^\prime = k_\perp^\prime/a$.
For the odd solution, the momentum perpendicular to the well is quantized by the condition 
\begin{multline}
\zeta_n \cot(\zeta_n) = -\sqrt{\frac{2\mathcal{W}a}{\hbar c} \sqrt{\zeta^2_n+\frac{m^2 c^2}{\hbar^2} a^2} - \zeta^2_n - \frac{\mathcal{W}^2 a^2}{\hbar^2 c^2}} 
\\
= -\zeta^\prime_n
\end{multline}
where $\zeta = k_\perp /a$ and $\zeta^\prime = k_\perp^\prime/a$.
The allowed energies are given by
\begin{eqnarray}
E_n &=& -W + \sqrt{m^2 c^4 + \hbar^2 c^2 (k^2_\parallel + \zeta_n^2/a^2)} 
\\
&=& \sqrt{m^2 c^4 + \hbar^2 c^2 (k^2_\parallel - {\zeta^\prime_n}^2/a^2 )}.
\end{eqnarray}
where $E_n \geq mc^2 - \mathcal{W}$.



\begin{thebibliography}{99}

\bibitem{1}
 R. Dick, \textit{Advanced Quantum Mechanics: Materials and Photons}
(Springer, New York, 2012). 

\bibitem{2}
R. Dick, Dimensional Effects on Densities of States and Interactions
in Nanostructures, Nanoscale Res. Lett. \textbf{5}, 1546 (2010).

\bibitem{3}
A. C. Zulkoskey, R. Dick, and K. Tanaka, Inter-dimensional
Effects in Systems with Quasi-Relativistic Dispersion Relations, Phys.
Rev. A \textbf{89}, 052103 (2014).

\bibitem{4}
A. H. Catro Neto, F. Guinea, N. M. R. Peres, K. S. Novoselov, and A. K. Geim, 
The electronic properties of graphene, Rev. Mod. Phys. \textbf{81}, 109 (2009).

\bibitem{5}
S. M. Young, S. Zaheer, J. C. Y. Teo, C. L. Kane, E. J. Mele, and
A. M. Rappe, Dirac Semimetal in Three Dimensions, Phys. Rev. Lett.
\textbf{108}, 140405 (2012).

\bibitem{6}
S. M. Young and C. L. Kane, Dirac Semimetals in Two Dimensions, 
Phys. Rev. Lett. \textbf{115}, 126803 (2015).

\bibitem{7}
J. E. Moore, The Birth of Topological Insulators, Nature \textbf{464}, 194 (2010).

\bibitem{8}
M. Z. Hasan, C. L. Kane, Topological Insulators, Rev. Mod. Phys. \textbf{82}, 3045 (2010).

\bibitem{9}
S. Borisenko, Q. Gibson, D. Evtushinsky, V. Zabolotnyy, 
B. Büchner, and R. J. Cava, Experimental Realization of a 
Three-Dimensional Dirac Semimetal, Phys. Rev. Lett. \textbf{113}, 027603 (2014).

\bibitem{10}
Z. K. Liu, J. Jiang, B. Zhou, Z. J. Wang, Y. Zhang, H. M. Weng, D. Prabhakaran, 
S-K. Mo, H. Peng, P. Dudin, T. Kim, M. Hoesch, Z. Fang, X. Dai, Z. X. Shen, 
D. L. Feng, Z. Hussain,
and Y. L. Chen, A Stable Three-dimensional Topological Dirac Semimetal Cd3As2, Nat. Mater. \textbf{13}, 677 (2014).

\bibitem{11}
T. O. Wehlinga, A. M. Black-Schafferc, and A. V. Balatskyd, Dirac Materials, Adv. Phys. \textbf{76}, 1 (2014).

\bibitem{12} 
K. A. Sveshnikov and P. K. Silaev, Quasiexact Solution of
a Relativistic Finite-Difference Analogue of the Schr\"{o}dinger Equation
for a Rectangular Potential Well, Theor. Math. Phys.
\textbf{132}, 1242 (2002); Quasiexact Solution of the Problem of 
Relativistic Bound States in the $(1+1)$-Dimensional Case,
Theor. Math. Phys. \textbf{149}, 1665 (2006).

\bibitem{13} 
V. Yu. Ananchenko, A. V. Sushchevskii, M. Sh. Pevzner, and
D. V. Kholod, Relativistic Fermions in a Spherically Symmetric Potential
Well of Finite Depth in a Two-Dimensional Space, Russian Physics Journal
\textbf{54}, 1256 (2011).


\end{thebibliography}
\end{document}